# Illustration of spin-half wave-functions, and other fractional-spin structures

Zeev Burshtein*
Department of Materials Engineering, Ben-Gurion University of the Negev, P.O. Box 653, Beer-Sheva 84105, Israel
*email address: zeevb@bgu.ac.il
Phone: +972-52-838-9884

**Abstract**
We provide an illustration of a geometric structure that exhibits the spin-half wave-function characteristics, namely that its full rotation displays a phase inversion. We show that general fractional-spin geometric structures are possible, along with sketchy discussions concerning group-theoretical and quantum descriptions, as well as pose some queries. Synthesis of a $(PF_3)_n$-chain polymer in the form of a twisted ring is proposed as a realization of a quazi-spin-1/3 structure.

## 1. Introduction

The occurrence in nature of Fermions exhibiting the spin=1/2 property, while being well established, carries with it a sense of mystery. Especially difficult to visualize and comprehend was the stated fact that a "full rotation of the particle wave-function brings about a phase inversion"; or alternatively, that "two full rotations of the particle around the Z-axis are needed to make the wave function appear the same". In this communication we not only provide an illustration of a geometric structure that exhibits this particular trait, but also show that fractional spin geometric structures are possible, along with sketchy highlights towards group-theoretical and quantum descriptions.

In 1961, T. H. R. Skyrme published some inspiring theoretical observations, that field configurations of bosons which are topologically twisted may behave like fermions [1-5]. In 2010, R. M. Wiener [6] suggested that all known fermions are in fact skyrmions, namely topologically twisted coherent boson states (solitons); in simplistic terms, that all fermions are not elementary, but rather quasi-particles, say "compound molecules" in the realm of particle physics. It would also imply that particle fermionic effects are apparent only at large distances. The non-observation in nature of bosonic leptons and baryons was considered as a supporting example to the conjecture. These issues are not in the scope of the present paper, except for the purpose of linking the "twisting" term we use hereinafter to the broader field of particle physics and cosmology.

## 2. The spin-half wave-function

Let us first summarize some well known quantum mechanical properties of the spin-half wave-function. The Fermionic wave-function (spinor) may be written as a two-dimensional vector of single valued functions of space $\underline{r}$ and time t

$$u(\underline{r},t) = \begin{pmatrix} \psi_+(\underline{r},t) \\ \psi_-(\underline{r},t) \end{pmatrix} . \tag{1}$$

The eigenfunctions of the $\mathbf{s}_z$ and $\mathbf{s}^2$ spin-half angular momentum operators are

$$u_+^{(1/2)}(\underline{r}) = \begin{pmatrix} \psi_+(\underline{r}) \\ 0 \end{pmatrix} \quad ; \quad u_-^{(1/2)}(\underline{r}) = \begin{pmatrix} 0 \\ \psi_-(\underline{r}) \end{pmatrix} . \tag{2}$$

$u_+$ is the "spin-up" wave-function, and $u_-$ is the "spin-down" wave-function. The spin operators acting on these wave functions give

$$\mathbf{s}_z u_+ = \tfrac{1}{2}\hbar u_+ \; ; \quad \mathbf{s}_z u_- = -\tfrac{1}{2}\hbar u_- \; ; \quad \mathbf{s}^2 u_\pm = \tfrac{3}{4}\hbar^2 u_\pm \; , \tag{3}$$

where $\hbar$ is the reduced Planck constant. The operator $\mathbf{s}_z$ actions can be represented by the matrix

$$\sigma_z = \frac{\hbar}{2}\begin{pmatrix} 1 & 0 \\ 0 & -1 \end{pmatrix} . \tag{4}$$

The wave-function rotation by an angle $\alpha$ is given by the matrix

$$\Gamma(\alpha) = \begin{pmatrix} e^{i\alpha/2} & 0 \\ 0 & e^{-i\alpha/2} \end{pmatrix} . \tag{5}$$

Performing a full rotation on the spinor wave-function means operating on it with $\Gamma(2\pi)$:

$$\Gamma(2\pi)\begin{pmatrix}\psi_+(\mathbf{r},t)\\\psi_-(\mathbf{r},t)\end{pmatrix}=\begin{pmatrix}e^{i2\pi/2} & 0\\0 & e^{-i2\pi/2}\end{pmatrix}\begin{pmatrix}\psi_+(\mathbf{r},t)\\\psi_-(\mathbf{r},t)\end{pmatrix}$$
$$=\begin{pmatrix}-1 & 0\\0 & -1\end{pmatrix}\begin{pmatrix}\psi_+(\mathbf{r},t)\\\psi_-(\mathbf{r},t)\end{pmatrix}=-\begin{pmatrix}\psi_+(\mathbf{r},t)\\\psi_-(\mathbf{r},t)\end{pmatrix}.\quad(6)$$

The generation and destruction operators $\mathbf{b}^\dagger \equiv \hbar^{-1}(\mathbf{s}_x+i\mathbf{s}_y)$ and $\mathbf{b}\equiv\hbar^{-1}(\mathbf{s}_x-i\mathbf{s}_y)$, respectively, where $\mathbf{s}_x$ and $\mathbf{s}_y$ are the x-, and y-components of the angular momentum operator, respectively, satisfy the anti-commutation relation

$$\{\mathbf{b}^\dagger,\mathbf{b}\}\equiv\mathbf{b}^\dagger\mathbf{b}+\mathbf{b}^\dagger\mathbf{b}=1\quad.\quad(7)$$

This property is in fact a key-step to proving that the only allowed eigen values of $\mathbf{s}_z$ are $\pm(1/2)\hbar$ (Eq. (3) above), which was the key consideration leading to Pauli's exclusion principle, hence the Fermi-Dirac statistics.

By Eq. (6), a full rotation brought about a phase inversion of the wave function. For visualizing this seemingly incomprehensible spin-half property, one needs to realize that the wave-function *should be observed in the laboratory frame*; namely that the phase angle evolution as the particle rotates must be viewed at a fixed point of the lab frame. Recall that this is the foundation of the general relations between infinitesimal rotations and angular momentum operators. For the $\mathbf{J}_z$ operator, the relation is expressed as

$$\mathbf{J}_z\equiv-i\hbar\frac{\partial}{\partial\varphi}\quad,\quad(8)$$

where $\varphi$ is the rotation angle of the rigid wave-function. The derivative operator on the right-hand side of (8) performs a change in the particle wave-function by rotating it rigidly around the Z-axis while observing it in the lab frame. On the left-hand side, the angular momentum operator $\mathbf{J}_z$ operates on the particle wave-function in the particle rest frame. The identity between the two sides indicates that the results of these two different operations are the same. Further recall, that an expression for a finite rotation operation $\Gamma(\alpha)$ is derived by performing a numerous succession of infinitesimal rotations per the right-hand side of Eq. (8). The expression for $\Gamma(\alpha)$ in Eq. (5) is obtained for the specific spin-1/2 case.

In Fig. 1 we show a geometric structure that exhibits a spin-half topology. The figure describes a Moebius stripe wound around the Z axis. The arrows marked across the stripe describe the wave function's phase belonging to the points along the central circle; namely the projection of each arrow on the Z axis indicates, for example, the real part of the function, while its projection on the (X,Y) plane correspondingly indicates the imaginary part. For clarity, the arrows are also given a dark "back" and a bright "front".

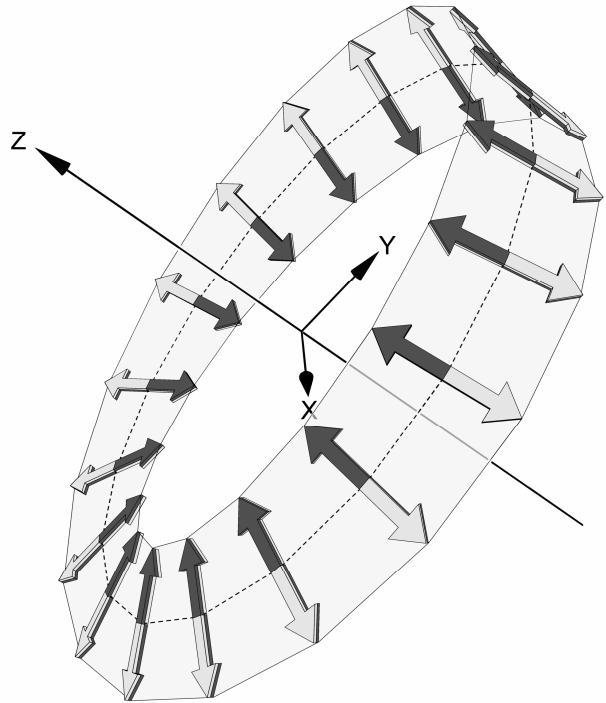

**Fig. 1** Demonstration of the internal structure of a physical system, whose rotational symmetry exhibits a spin-half feature; namely, as viewed in the lab frame it has to complete two full rotations around the Z-axis to have its phase return to its original posture.

To provide some vivid understanding of the lab-frame concept, imagine a reporter settling itself at a certain point on the central circle, *but never moves even though the structure rotates rigidly about the Z axis.* The reporter's duty is to record the situation evolving at his specific location. As viewed by the reporter, different arrow pairs successively appear *at his site*, representing the gradually changing phase. It is easily verified, that a full rotation will present the reporter with an arrow-pair that inverted both their directions and fronts relative to the Z axis. From his point of view, only an additional full rotation will bring back at his site an arrow-pair identical with the original. Now for the spin-half wave-function, the above property *is shared by all points of space*.

Notably, it was the requirement of viewing the system in the lab frame that allowed noticing the phase inversion property upon a full rotation of the structure shown in Fig. 1.

Actually, upon a full rotation, the structure returns to its original posture, as is fundamentally true for all rigid geometric structures!

We recall now, that a Moebius stripe is obtained by acting to fold a simple stripe to form a cylindrical ring, yet on joining the two edges, the two edges are oppositely twisted by $180°$. Fig. 1's structure describes the $u_+^{(1/2)}$ wave-function, namely the phase evolution upon rotation of the $\psi_+$ component (Eqs. (5) and (6)). A description of the $u_-^{(1/2)}$ wave function will be obtained by *twisting the stripe edges in the opposite sense* when joined to form the Moebius stripe, or alternatively by letting the stripe rotate in the opposite sense. In fact, the group-theoretical approach of forming a double-group to describe spin-half structures simulates the Moebius stripe topology: a double-group is an extension of a symmetry group, where per each symmetry rotation operation $C(\alpha)$ by an angle $\alpha$ about a symmetry axis belonging to the group, a rotation $C(\alpha + 2\pi)$ is added [7,8]. While the identity operation E (angles $\alpha = 4\pi n$, where n is an integer including zero) is represented by an identity matrix I, the related "twisted identity" operation $\hat{E}$ (angles $\alpha = 4\pi n + 2\pi$) is represented by the negative identity matrix $-I$; Namely, $\Gamma(E) = I$ and $\Gamma(\hat{E}) = -I$.

## 3. Structures of spin-1/3 topology

Following the same logic that brought about the formation of Fig. 1's structure, in Fig. 2 we present a phase structure that exhibits a spin-1/3 topology. The figure describes a chain of triplet-vector units wound as a ring around the Z axis. All angles between adjacent vectors within each unit are $2\pi/3$. Each arrow describes the phase of a wave function component belonging to the points along the central circle; namely the projection of each arrow on the Z axis indicates, for example, the real part of the component function, while its projection on the (X,Y) plane correspondingly indicates the imaginary part. For clarity, the arrows are also given a dark "back" and a bright "front". Triplet-vectors of adjacent units are twisted such that upon a full length of the ring chain, the accumulated twist is $2\pi/3$. It is thus trivially verified that a full ring rotation will present a lab-frame observer with an arrow-triplet rotated by $2\pi/3$. The performance of three full rotations is required to bring back to the lab-frame observer an arrow-triplet identical with the original. In that respect, it is a spin-one-third structure! We consider the above property to belong to *any point in space* for a spin-one-third functional.

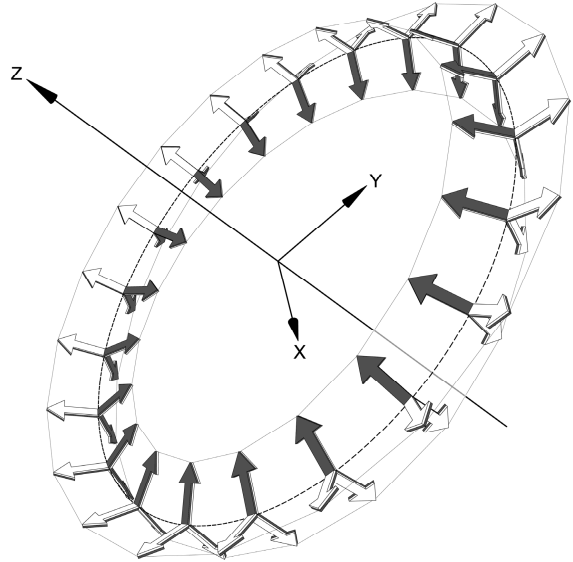

**Fig. 2** Demonstration of the internal structure of a system, whose rotational symmetry exhibits a spin-one-third feature; namely, as viewed in the lab frame it has to complete three full rotations around the Z-axis to have its phase return to its original posture.

If the structure is twisting in the opposite direction, or alternatively the ring rotates around the Z-axis in the opposite sense, a spin-**minus**-one-third obtains. Notably, the spin-one-third structure of Fig. 2 exhibits two independent states: plus and minus (or "up" and "down"), same as the spin-half structure.

For a group-theoretical description of the spin-1/3 topology, one may consider the forming of a triple-group. Namely, per each symmetric rotation operation $C(\alpha)$ by an angle $\alpha$ about a symmetry axis belonging to the group, two rotation operation would be added, $C(\alpha + 2\pi)$ and $C(\alpha + 4\pi)$. The identity operation E (angles $\alpha = 6\pi n$, where $n$ is an integer including zero) would be represented by an identity matrix I; a related "first twisted identity" operation $\hat{E}$ (angles $\alpha = 6\pi n + 2\pi$) would be represented by a "twisted identity matrix" $e^{2\pi i/3}I$; a related "second twisted identity" operation $\hat{\hat{E}}$ (angles $\alpha = 6\pi n + 4\pi$) would be represented by a "second twisted identity matrix" $e^{4\pi i/3}I$. Namely, $\Gamma(E)=I$, $\Gamma(\hat{E})=e^{2\pi i/3}I$, and $\Gamma(\hat{\hat{E}})=e^{4\pi i/3}I$.

On the formal quantum mechanical side, some features of the spin-half may be quite simply extended to the spin-one-third. As extension of Eq. (1), the wave-function may be written as a 2×3 matrix of single valued functions of space $\underline{r}$ and time t

$$u^{(1/3)}(\underline{r},t) = \begin{pmatrix} \psi_+(\underline{r},t), & \psi_+(\underline{r},t)e^{2\pi i/3}, & \psi_+(\underline{r},t)e^{4\pi i/3} \\ \psi_-(\underline{r},t), & \psi_-(\underline{r},t)e^{-2\pi i/3}, & \psi_-(\underline{r},t)e^{-4\pi i/3} \end{pmatrix}. \tag{9}$$

Should proper quantum operators be constructed, their "spin-up" $u_+$ and "spin-down" $u_-$ eigen-functions may be expected to be (see Eq. (2))

$$u_+^{(1/3)}(\underline{r}) = \begin{pmatrix} \psi_+(\underline{r}), & \psi_+(\underline{r})e^{2\pi i/3}, & \psi_+(\underline{r})e^{4\pi i/3} \\ 0, & 0, & 0 \end{pmatrix};$$

$$u_-^{(1/3)}(\underline{r}) = \begin{pmatrix} 0, & 0, & 0 \\ \psi_-(\underline{r}), & \psi_-(\underline{r})e^{-2\pi i/3}, & \psi_-(\underline{r})e^{-4\pi i/3} \end{pmatrix}. \tag{10}$$

The wave-function rotation by an angle $\alpha$ is given by the matrix (see Eq. (5)).

$$\Gamma(\alpha) = \begin{pmatrix} e^{i\alpha/3} & 0 \\ 0 & e^{-i\alpha/3} \end{pmatrix}. \tag{11}$$

Performing a full rotation on the wave-function means operating on it with $\Gamma(2\pi)$ (see Eq. (6)):

$$\Gamma(2\pi)u^{(1/3)}(\underline{r},t) = \begin{pmatrix} e^{i2\pi/3} & 0 \\ 0 & e^{-i2\pi/3} \end{pmatrix} \begin{pmatrix} \psi_+(\underline{r},t), & \psi_+(\underline{r},t)e^{2\pi i/3}, & \psi_+(\underline{r},t)e^{4\pi i/3} \\ \psi_-(\underline{r},t), & \psi_-(\underline{r},t)e^{-2\pi i/3}, & \psi_-(\underline{r},t)e^{-4\pi i/3} \end{pmatrix}$$
$$= \begin{pmatrix} \psi_+(\underline{r},t)e^{2\pi i/3}, & \psi_+(\underline{r},t)e^{4\pi i/3}, & \psi_+(\underline{r},t) \\ \psi_-(\underline{r},t)e^{-2\pi i/3}, & \psi_-(\underline{r},t)e^{-4\pi i/3}, & \psi_-(\underline{r},t) \end{pmatrix} . \tag{12}$$

Thus a full rotation brings about a cyclic transformation among the components of each spin-up and spin-down wave-functions. Obviously, three full rotations $\Gamma(6\pi)$ return the $u^{(1/3)}(\underline{r},t)$ function to its original form.

Naturally, one wonders whether spin-one-third particles or quasi-particles may occur in nature; and if they do, what are their eigen-values in relation to angular momentum, and would they be bosons, fermions, or exhibit more complex statistics. Some considerations will be presented in Section 4 below; a comprehensive study is yet to be performed separately. However, in order not to leave this query on philosophical grounds solely, in Fig. 3 we present a geometric "molecular" structure that really exhibits a quasi-spin-1/3 topology! The structure appears as an axial long polymer of central, type-I atoms, each forming "equatorial" three-fold rotational symmetry bonds perpendicular to the chain with type II atoms; then this long chain is wound around the Z-axis to form a closed ring, however with a "twist" that bonds each type-II chain atom to the $2\pi/3$ rotated chain atom of the other edge. Observation of the structure on it's equator in the lab frame readily reveals that three full rotations around the Z-axis are needed to have the structure return to a posture identical with the one it had at the beginning of rotation. A naive chemical consideration suggests that synthesizing such molecule with the penta-valence phosphorus atom P as the equatorial, type-I atom, and the singly-valence fluorine atom as the type-II atom, is possible. In other words, we propose the synthesis of a $(PF_3)_n$-chain polymer in the form of a twisted ring as a realization of a quazi-1/3-spin structure. Other combinations may probably be equally suitable or even better.

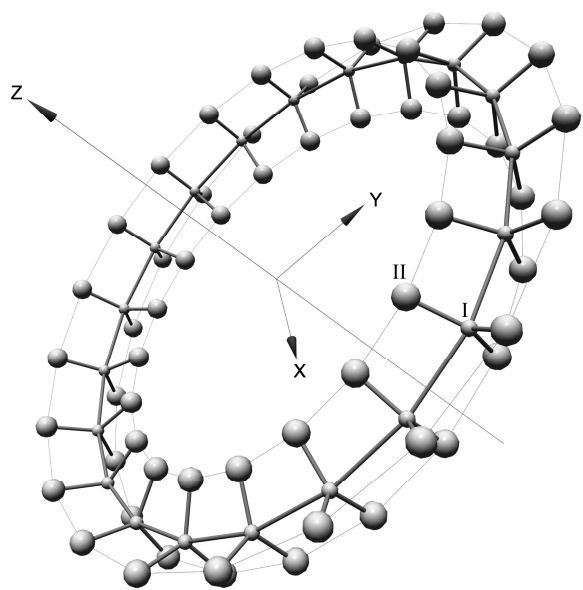

**Fig. 3** A model of a geometric "molecular" structure that exhibits a quasi-spin-1/3 topology. The central ring is a chain composed of n, type-I atoms, forming equatorial 3-fold rotational symmetry bonds perpendicular to the chain with type-II atoms. A gradual twist of the 3-fold groups exhibits a $2\pi/3$ angular twist per n steps.

The above statement that the structure exhibits "a quasi-spin-1/3 topology", and not simply a "spin-1/3 topology", relates to the fact that only points on the chain-equator exhibit that trait, and not the entire universe. Still, a study of the electronic states of such molecular structures should be very helpful. Consider an electron belonging to a molecular orbital revolving coherently about the Z-axis along the type-I atoms. The electron must experience a phase change during motion, which should simulate effects related to the spin-1/3 reality. An experimental study of such molecules may reveal the allowed angular momentum values and their related magnetic moments.

**4. Structures of general, partial spin topology**

One easily realizes that a variety of partial spin structures may be visualized by simply winding, twisting, and end-joining of an originally symmetric linear chain. In fact, partial $1/p$ spin structures for any $p > 1$ may be considered. It is highly desirable that such structures be studied both theoretically and experimentally. Considering electrons for example, the fundamental questions that arise are: what would be the allowed angular momentum values for a spinless electron revolving coherently about the Z axis, experiencing fractional phase-changes during revolution; what would be the angular moments occupation statistics; and what would be the magnetic moment induced per angular momentum quantum.

As is well known [9], the only allowed angular momentum eigen-states of both $\mathbf{J}_Z$ and $\mathbf{J}^2$ angular momentum operators are those belonging to the quantum numbers $j = 0, \frac{1}{2}, 1, \frac{3}{2}, 2, \frac{5}{2}, \cdots$. Within that scheme, no 1/3 or a general fractional spin may be allowed. It is thus our suggestion that the constraint of being an eigen-state of $\mathbf{J}^2$ be lifted; namely, the only requirement would be being an eigen-state of $\mathbf{J}_Z$. We further propose that the eigen-value would be $(1/p)$, namely $\mathbf{J}_Z^{(1/p)} u_+^{(1/p)} = (1/p)\hbar u_+^{(1/p)}$ and $\mathbf{J}_Z^{(1/p)} u_-^{(1/p)} = -(1/p)\hbar u_-^{(1/p)}$. Lifting of the total angular momentum conservation condition is obviously very bothersome. For molecular systems such as in Fig. 3 one may justify its application to electrons on grounds that the total angular momentum is conserved by the entire heavy system that includes the electrons and especially the heavy molecular nucleonic skeleton.

**5. Summary**

To alleviate the sense of mystery entailing the spin-half particle property, which states that a full rotation of the particle wave-function brings about a phase inversion, we provide an illustration of a geometric structure exhibiting this trait. It is based on the symmetry of a Moebius stripe, which is known to be obtained by folding a simple stripe to form a cylindrical ring, yet on joining the two edges, the two edges are twisted by 180°. We then show that fractional-spin geometric structures are possible. They are obtained by administering a $2\pi/p$ twist on joining the ends of a linear structure of originally p-fold axial symmetry wound to form a ring. An example of a particular quazi-1/3-spin structure is described in detail. Synthesis of a $(PF_3)_n$-chain polymer in the form of a twisted ring is proposed as a realization of a quazi-1/3-spin structure. Some queries as well as sketchy highlights towards group-theoretical and quantum descriptions are discussed.

## 6. Acknowledgment

I am indebted to my son, architect Joseph Burshtein, for drawing the figures.